\documentclass[pra,aps]{revtex4}
\usepackage{graphicx,amsfonts,bm,amsmath}

\newcommand{\rl}{\rangle\!\langle}

\begin{document}
\title{Four-wave mixing spectroscopy of quantum dot molecules}
\author{Anna Sitek and Pawe{\l} Machnikowski}
 \affiliation{Institute of Physics, Wroc{\l}aw University of
Technology, 50-370 Wroc{\l}aw, Poland}

\begin{abstract}
We study theoretically the nonlinear four-wave mixing response of an
ensemble of coupled pairs of quantum dots (quantum dot molecules). We
discuss the shape of the echo signal depending on the parameters
of the ensemble: the statistics of transition energies and the degree
of size correlations between the dots forming the molecules.
\end{abstract}


\maketitle

\section{Introduction}

Optical experiments on quantum dot molecules (QDMs)
\cite{borri03,ortner03,bardot05} have shown that
QDMs differ in many ways from single quantum dots (QDs).
In a recent work \cite{sitek07a} we have discussed 
the effects of collective interaction between closely spaced dots  and 
the modes of electromagnetic field
(sub- and superradiance \cite{scheibner07}), as
manifested in the luminescence decay from a single QDM. 
Here we will consider the nonlinear four-wave mixing (FWM)
response from an ensemble of QDMs \cite{borri03}. 

The FWM spectroscopy is a powerful tool for extracting
information on lifetimes and homogeneous dephasing from inhomogeneous
ensembles \cite{borri99,borri01}. In the case of Markovian
decoherence, the interpretation of the measured FWM signal is
straightforward \cite{schafer02}. More care must be taken for
non-Markovian pure dephasing, where the FWM response does not
reproduce the optical response of a single system
\cite{vagov03}. As we shall see, the time evolution of the FWM
signal from QDMs is much more complicated than that from QDs and
carries information on the interaction between the dots and
on the correlations of their size. Therefore, before one can proceed
to the discussion of decoherence effects, it is necessary to analyze
all the factors that affect the shape of the FWM
response. This is the goal of the present paper. We will study the
shape of the FWM signal for a pair of QDs and its dependence on the
size distribution and size correlations between the dots, as well as
on the interaction between them, in absence of any decoherence.

\section{The model}

Each QDM will be modelled as a
four-level system with the basis states
$|00\rangle,|01\rangle,|10\rangle,|11\rangle$, corresponding to the
ground state (empty dots), an exciton in the second and first QD, 
and excitons in both QDs, respectively.
A FWM experiment consists in exciting an ensemble of QDMs with two
ultrashort laser pulses, arriving at $t_{1}=-\tau$ and $t_{2}=0$. We
assume that the pulses are spectrally very broad to assure resonance
with all the QDs in the ensemble. First, consider a single QDM
composed of two QDs with energies $E_{1,2}=E\pm\Delta$.
We will describe its evolution in a frame rotating with the frequency $E/\hbar$.
Then, in the rotating wave approximation, the Hamiltonian for this single
QDM is
\begin{eqnarray}
\label{ham}
H & = & \Delta\left( 
|1\rl 1|\otimes \mathbb{I} - \mathbb{I}\otimes|1\rl 1|\right)
+V\left(  |01\rl|10| + |10\rl|01| \right)\\
\nonumber
&&+\frac{1}{2}\sum_{i}f_{i}(t)\left[ 
e^{-i(\phi_{i}+Et_{i})}
\left( |0\rl 1|\otimes\mathbb{I}+\mathbb{I}\otimes|0\rl 1| \right) 
+\mathrm{H.c.} \right],
\end{eqnarray}
where $f_{i}$ and $\phi_{i}$ are the amplitude envelopes and phases of the
pulses, respectively, $V$ is the coupling between the dots (tunnel or
F{\"o}rster), $\mathbb{I}$ is the
unit operator and the tensor product decomposition refers to the two
QDs making up the molecule. The pulses do not overlap in time.

\section{The system evolution and FWM response}

If the durations of the pulses are much shorter than both
$\hbar/\Delta$ and $\hbar/V$, the action of each of them corresponds
to the independent rotation of the state of each QD, that is, to the
unitary transformation $\mathsf{U_{i}}=U_{i}\otimes U_{i}$, where 
$U_{i}=\cos(\alpha_{i}/2)\mathbb{I}
-i\sin(\alpha_{i}/2)[ e^{-i(\phi_{i}+Et_{i})}|0\rl 1|
+\mathrm{H.c.}]$
and $\alpha_{i}=\int_{-\infty}^{\infty}dtf_{i}(t)$ is the pulse area.
Between the pulses and after the second pulse the system evolution is
generated by the time-independent Hamiltonian given by the first two
terms in Eq.~(\ref{ham}). The resulting evolution operator reads
\begin{eqnarray*}
\mathsf{W}(t)&=&|00\rl 00|+|11\rl 11|+\cos(\Delta t)(|01\rl 01|+|10\rl 10|)\\
&&+i\sin(\Delta t)\left[  \cos(2\theta)(|01\rl 01|-|10\rl 10|)
-\sin(2\theta)(|01\rl 10|+|10\rl 01|) \right],
\end{eqnarray*}
where $\sin(2\theta)=V/\sqrt{V^{2}+\Delta^{2}}$.

The system state at a time $t>0$ is represented by the density matrix
\begin{displaymath}
\rho(t)=\mathsf{W}(t)\mathsf{U}_{2}\mathsf{W}(\tau)\mathsf{U}_{1}
\rho(0)[\mathsf{W}(t)\mathsf{U}_{2}\mathsf{W}(\tau)\mathsf{U}_{1}]^{\dag},
\end{displaymath}
where $\rho(0)=|00\rl 00|$ is the initial state. The optical
polarization of a single QDM under consideration is proportional to
$P(t)=\langle 11|\rho(t)(|01\rangle+|10\rangle)
+(\langle 01|+\langle 10|)\rho(t)|00\rangle+\mathrm{c.c.}$. In order
to extract the FWM polarization 
we pick out the terms containing the phase factor
$e^{i(2\phi_{2}-\phi_{1})}$. 
The total optical response from the sample is obtained by summing up
the contributions from individual QDMs with a weight factor
$g(E_{1},E_{2})$ reflecting the distribution of transition energies in
the sample (we neglect a possible variation of dipole moments). As a
result we obtain the third-order FWM signal
$P_{\mathrm{FWM}}(t)=\sum_{i}P_{i}(t)+\mathrm{c.c.}$, where
\begin{subequations}
\begin{eqnarray}\label{P}
P_{1}(t) & = & i\sin\alpha_{1}\sin^{2}\frac{\alpha_{2}}{2}
\int dE_{1}dE_{2} g(E_{1},E_{2})e^{-iE(t-\tau)}
\cos\left[ \Omega(t-\tau) \right], \\
P_{2}(t) & = & \frac{V}{4\Omega}\sin2\alpha_{1}\sin^{2}\alpha_{2}
\int dE_{1}dE_{2} g(E_{1},E_{2})e^{-iE(t-\tau)}
\sin\left[ \Omega(t-\tau) \right], \\
P_{3}(t) & = &
\frac{V}{2\Omega}\sin2\alpha_{1}\sin^{2}\frac{\alpha_{2}}{2}\cos\alpha_{2}
\int dE_{1}dE_{2} g(E_{1},E_{2})e^{-iE(t-\tau)}
\sin\left[ \Omega(t+\tau) \right],\\
\nonumber 
P_{4}(t) & = &
i\frac{2V^{2}}{\Omega^{2}}\sin\alpha_{1}
\sin^{2}\frac{\alpha_{2}}{2}\cos\alpha_{2} 
\in						t dE_{1}dE_{2} g(E_{1},E_{2})e^{-iE(t-\tau)}\\
&&\times\frac{\cos\left[ \Omega(t-\tau) \right]
+\cos\left[ \Omega(t+\tau)\right] }{2}, 
\end{eqnarray}
\end{subequations}
where $\Omega=\sqrt{V^{2}+\Delta^{2}}$. 
We assume a Gaussian distribution function
\begin{eqnarray}\label{g}
\lefteqn{g(E_{1},E_{2})=}\\
\nonumber
&&\frac{1}{2\pi\sigma^{2}\sqrt{1-\rho^{2}}}
\exp\left[  -\frac{\left(E_{1}-\bar{E}_{1}\right)^{2}
-2\rho\left(E_{1}-\bar{E}_{1}\right)\left(E_{2}-\bar{E}_{2}\right)
+\left(E_{2}-\bar{E}_{2}\right)^{2}}{2\left(1-\rho^{2}\right)\sigma^{2}}
\right]
\end{eqnarray}
with identical energy variances for both QDs (consistent with the
symmetric photoluminescence spectrum observed in the experiment
\cite{borri03}) and a correlation coefficient $\rho$. Note that this
distribution corresponds to an uncorrelated Gaussian distribution of
the parameters $E$ and $\Delta$ with variances 
$\sigma_{E}=\sigma\sqrt{(1+\rho)}/\sqrt{2}$
and $\sigma_{\Delta}=\sigma\sqrt{(1-\rho)}/\sqrt{2}$ (a small variance
of the energy difference $\Delta$ implies correlated energies).
For the calculations we will use the values 
for the energy variance $\sigma=8$~meV and for the average energy
mismatch $\bar{\Delta}=(\bar{E}_{1}-\bar{E}_{2})/2=11$~meV
\cite{borri03}. In order to fix the relative amplitudes of the
contributions $P_{i}(t)$ we assume $\alpha_{i}\ll
1$ and expand the trigonometric coefficients in Eqs.~(\ref{P}-d).

The detection of weak signals originating from QDs is based
on a heterodyne technique \cite{borri99}: The response
$P_{\mathrm{FWM}}$ is superposed onto a reference pulse
$E_{\mathrm{ref}}(t)
=f_{\mathrm{ref}}(t-t_{0})e^{-i\overline{E}(t-t_{0})}+\mathrm{c.c.}$,
arriving at a time $t_{0}$. We assume 
a Gaussian envelope
$f_{\mathrm{ref}}(t)= 
\exp[-(1/2)(t/\tau_{\mathrm{ref}})^{2}]/(\sqrt{2\pi}\tau_{\mathrm{ref}})$.
The measured signal is proportional to
\begin{equation}\label{F}
F(t_{0},\tau)=\left| \int dt 
P_{\mathrm{FWM}}^{(+)}(t)E_{\mathrm{ref}}^{(-)}(t)\right|,
\end{equation}
where $P_{\mathrm{FWM}}^{(+)}$ and $E_{\mathrm{ref}}^{(-)}$ are the
positive frequency part of the FWM signal and the negative frequency
part of the reference pulse, respectively.

\begin{figure}[tb]
\begin{center}
\unitlength 1mm
\begin{picture}(90,45)(0,5)
\put(0,0){\resizebox{90mm}{!}{\includegraphics{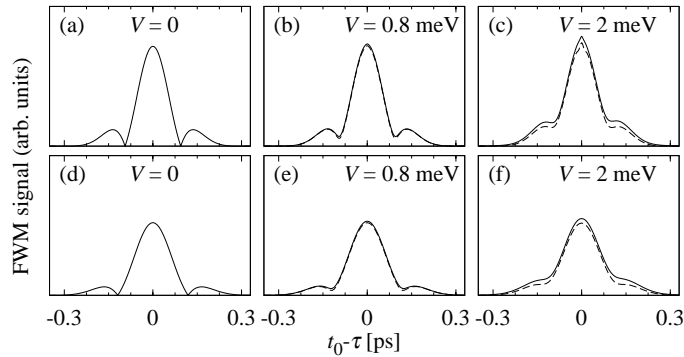}}}
\end{picture}
\end{center}
\caption{FWM signal from the QDM sample detected with an
ultrashort  reference pulse (a-c) or with a 100 fs FWHM pulse
(d-f). Solid lines: $\rho=0$; dashed lines: $\rho=0.875$.}
\label{fig:tr}
\end{figure}

In the case of decoupled QDs
($V=0$), the only non-vanishing contribution to
$P_{\mathrm{FWM}}$ is $P_{1}$. Inserting Eq.~(\ref{g}) into
Eq.~(\ref{P}) and using Eq.~(\ref{F}) one finds 
\begin{displaymath}
F(t_{0},\tau)=\frac{1}{\sqrt{\tau_{\mathrm{ref}}^{2}\sigma^{2}+1}}
\exp\left[ 
-\frac{\sigma^{2}(\tau-t_{0})^{2}+\tau_{\mathrm{ref}}^{2}
\bar{\Delta}^{2}}{2(\tau_{\mathrm{ref}}^{2}\sigma^{2}+1)}
\right] 
\cos\frac{\bar{\Delta}(\tau-t_{0})}{\tau_{\mathrm{ref}}^{2}\sigma^{2}+1}.
\end{displaymath}
This result is plotted in Fig.~\ref{fig:tr}a,d for 
an ultrashort reference pulse and for a pulse with a realistic
duration of $\tau_{\mathrm{ref}}=43$~fs, i.e., 100~fs full width at
half maximum (FWHM).
The obtained result is remarkable for two reasons. First, the shape of the
echo peak for noninteracting dots does not depend on the correlations
between the transition energies $E_{1}$ and $E_{2}$. Second, the
beats related to the energy mismatch are always in phase with the FWM
echo, so that they do not lead to oscillations in the time-integrated
signal. The only trace of these beats are the oscillations in the
tails of the echo peak. Third, the amplitude of the echo contains the
factor
$\exp[-(1/2)\tau_{\mathrm{ref}}^{2}\bar{\Delta}^{2}/
(\tau_{\mathrm{ref}}^{2}\sigma^{2}+1)]$ and becomes small if the
QDM is formed by very different dots.

If the QDs are coupled, the other three terms contribute
to the signal. This results in a modification of the shape of the echo
peak, which now depends on the degree of correlation between the
energy parameters of the two dots. In Fig.~\ref{fig:tr} we show 
the results obtained by numerical integration. 
For $V=0.8$~meV (Fig.~\ref{fig:tr}b,e), which is a realistic value of the
F{\"o}rster coupling for $5$~nm distance between the dots, the shape
is only slightly modified. A much larger effect is observed for
stronger coupling (as might result from tunneling). 
For the parameters chosen here, the characteristic sharp
feature appearing at the top of the response peak for strong couplings
is considerably smeared out by
the 100 fs reference pulse (Fig.~\ref{fig:tr}c,f).
Nevertheless, the evolution of the echo shape with growing coupling
remains clear.

\begin{figure}[tb]
\begin{center}
\unitlength 1mm
\begin{picture}(70,30)(0,5)
\put(0,0){\resizebox{70mm}{!}{\includegraphics{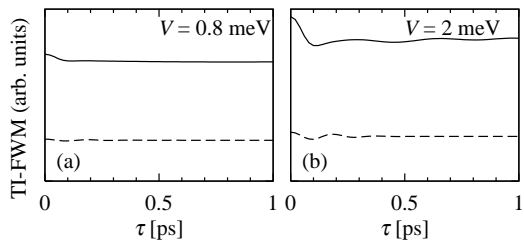}}}
\end{picture}
\end{center}
\caption{Time-integrated FWM signal as a function of the delay for two
values of the coupling as shown and for $\rho=0$ (solid) and
$\rho=0.875$ (dashed).}
\label{fig:ti}
\end{figure}

As mentioned above, in the limit of vanishing interaction there are no
beats in the time-integrated FWM signal as a function of the
delay $\tau$. This changes for $V\neq 0$ due to the contribution from
the term $P_{3}(t)$ and second part of $P_{4}(t)$ (Fig.~\ref{fig:ti}). 
At the center of
the echo peak, $t=\tau$, these terms have the variable phase factors
$\cos 2\Omega \tau$ and $\sin 2\Omega\tau$. As a result, they lead to
oscillations of the integrated FWM signal as a function of $\tau$ but
only as long as $\tau\sigma_{\Delta}\lesssim 1$. For longer times these
terms do not contribute because of the phase averaging due to 
inhomogeneous distribution of energy differences $\Delta$. Such
oscillations of the time-integrated signal might be a useful signature
of interaction between the QDs. Unfortunately, for the system
parameters used here they are present only for $\tau\lesssim 1 $~ps
and will be covered by phonon-related features \cite{borri03}.

\section{Conclusion}

Our results show that the shape of the time-resolved 
FWM signal may provide some clue on the properties of the QDMs
in the ensemble, including coupling and, for coupled dots, 
size correlations between the
QDs. We have studied only the amplitude of the FWM response. It should
be noted, however, that the terms $P_{2}$ and $P_{3}$ are real, while
the other two are purely imaginary. Therefore, much more information
is contained in the phase properties of the signal. Moreover,
nonlinear experiments can be performed beyond the perturbative regime
\cite{borri02a}. This might allow one to selectively investigate
selected contributions to the FWM signal by exploiting their different
dependence on the pulse areas $\alpha_{1,2}$.

The authors are very grateful to W. Langbein for useful hints concerning the
experimental technique.



\end{document}